\documentstyle[aps,prb,twocolumn,floats,epsfig]{revtex}

\begin{document}

\bibliographystyle{prsty}

\title{ 
\begin{flushleft}
{\small \em submitted to}\\
{\small 
PHYSICAL REVIEW B 
\hfill
VOLUME XX, 
NUMBER X 
$\qquad$
\hfill 
MONTH XXX
}
\end{flushleft}  
Quantum statistical metastability for a finite spin
\vspace{-1mm}
}

\author{
D. A. Garanin \cite{e-gar}, 
}

\address{
Max-Planck-Institut f\"ur Physik komplexer Systeme, N\"othnitzer Strasse 38,
D-01187 Dresden, Germany\\ }

\author{
E. M. Chudnovsky \cite{e-chu} 
}

\address{
Department of Physics and Astronomy, City University of New York -- 
Lehman College, Bedford Park Boulevard West, Bronx, New York 10468-1589 \\
\smallskip
{\rm(Received 28 May 2000)}
\bigskip\\
\parbox{14.2cm}
{\rm
We study quantum-classical escape-rate transitions for uniaxial and biaxial models with finite spins $S=10$ (such as Mn$_{12}$Ac and Fe$_8$) and $S=100$ by a direct numerical approach.
At second-order transitions the level making a dominant contribution into thermally assisted tunneling changes gradually with temperature whereas at first-order transitions a group of levels is skipped.
For finite spins, the quasiclassical boundaries between first- and second-order transitions are shifted, favoring a second-order transition: For Fe$_8$ in zero field the transition should be first order according to a theory with $S\to\infty$, but we show that there are no skipped levels at the transition.
{\em Applying a field along the hard axis in Fe$_8$ makes transition the strongest first order.}
For the same model with $S=100$ we confirmed the existence of a region where a second-order transition is followed by a first-order transition [X. Mart\'\i nes Hidalgo and E. M. Chudnovsky, J. Phys.: Condensed Matter (in press)].
\smallskip
\begin{flushleft}
PACS numbers: 75.45.+j, 75.50.Tt
\end{flushleft}
} 
} 
\maketitle

\section{Introduction}
\label{sec_introduction}

Recent experimental discovery of spin tunneling in large-spin compounds such as Mn$_{12}$Ac and Fe$_8$ ($S=10$) also stimulated theoretical investigation of spin models with $S\gg 1$ which show different kinds of transition between the classical mechanism of thermal activation over the potential barrier $\Delta U$ at temperatures $T_0 < T \ll \Delta U$ and the quantum regimes involving tunneling under the barrier at $T < T_0$.
The quantum-classical transition temperature $T_0$ becomes well defined in the quasiclassical limit $S\to\infty$ and is of order $T_0\sim \Delta U/S$, where $\Delta U$ is the barrier height.
Possible types of the quantum-classical transition for a general model have been classified in Ref.\ \onlinecite{chu92}, the two main scenarios being the so-called first-order transition and the second-order transition.
At the second-order transition the energy $E^*$ with which the system is crossing the barrier begins to move down from the value $E_c$ corresponding to the top of the barrier, and $E^*(T)$ approaches the bottom of the well $E_{\rm min}$ at $T=0$.
That is, for $T<T_0$ there is a thermally assisted tunneling: Thermal activation up to the energy $E=E^*$ is followed by the tunneling at this energy level. 
At the first-order transition $E^*$ abruptly changes from $E_c$ to some lower value and then, again, $E^*(T)$ approaches the bottom of the well at $T=0$.
In this situation some interval of energy is skipped and it does not contibute to the escape from the metastable well at any temperature.
There are more exotic cases such as a second-order transition followed by a first-order transition.

Whereas for quantum particles in the well it is difficult to realize transitions other than the second order, it has been recently shown that there are all the types of quantum-classical transitions in the spin model with the Hamiltonian
\begin{equation}\label{sham}
{\cal H} = -DS_z^2 +BS_x^2 - H_xS_x - H_yS_y - H_zS_z, 
\end{equation}
which is convenient to parametrize in terms of the reduced hard-axis anisotropy $b\equiv B/D$ and the reduced fields $h_x\equiv H_x/(2SD)$, etc.
In Ref.\ \onlinecite{garchu97} it was shown that in the uniaxial model ($B=0$) transverse field controls the order of transition which is first order for small transverse fields.
In Ref.\ \onlinecite{chugar97} the exact quasiclassical value of the critical transverse field $h=h_c=1/4$ has been obtained.
In Ref.\ \onlinecite{garmarchu98} the whole phase diagram of escape-rate transitions for the uniaxial model in the plane $h_x,h_z$ was drawn, the boundary line $h_{xc}(h_z)$ going to zero at $h_z=1$.

For the biaxial model, $B\neq 0$, in zero field the transition was shown to be first order for $b<b_c=1$ and second order for $b>1$. \cite{liamueparzim98}
Longitudinal field suppresses the value of $b_c$, so that $b_c(h_z)$ vanishes at $h_z=1$ together with the potential barrier.\cite{garchu99prb}
The quasiclassical result in this case reads $b_c=(1-h_z^2)/(1+2h_z^2)$. \cite{zhaetal99,kim99} 

Field along the medium axis, $H_y$, in Eq.\ \ref{sham} also favors the second-order transition.
Some points of the boundary between the first- and second-order transitions have been obtained in Ref.\ \onlinecite{leemueparzim98} numerically, whereas the analytical expression for the boundary was obtained later in Ref.\ \onlinecite{kim99}.
Recently the phase diagram of the biaxial model with the fields along medium  ($H_y$) and easy ($H_z$) directions has been considered in Ref.\ \onlinecite{paryooyoo00}.  
The qualitative results are a combination of those of Ref.\ \onlinecite{garmarchu98} for the uniaxial model and those of Refs.\ \onlinecite{zhaetal99,kim99} for the biaxial model with the field $H_z$: Increasing of all $b$, $h_y$, and $h_z$ favors the second-order transition.

The most interesting model is the biaxial model with the field along the hard direction $H_x$, in which oscillations of the tunneling probability as function of $h_x$ have been established theoretically \cite{garg93} and experimentally.\cite{werses99}
The phase diagram of escape-rate transitions for this model has been recently considered in Refs.\ \onlinecite{chokim00,marchu00}.
Unlike all other models, there is a first-order transition for $h_x\sim b$, even for the large enough values of $h_x$ and $b$ which would alone cause a second-order transition.
That is, the order of transition can change as II-I-II with increasing $h_x$ or $b$.
Moreover, in Ref.\ \onlinecite{marchu00} ranges of parameters have been found where the second-order transition is followed by a first-order transition or a first-order transition is followed by another first-order transition with lowering temperature.

Thus the theoretical investigation of the escape-rate transitions in the spin system described by Eq.\ \ref{sham} in the quasiclassical approximation $S\gg 1$ is nearly completed.
Experimentally studied materials, however, have the moderate spin value $S=10$, which can result in deviations from the predictions of the quasiclassical theory.
Indeed, in Mn$_{12}$Ac in zero applied field one can expect a strong first-order transition but experiments of Kent {\em et al} \cite{kenetal00} show only one skipped energy level, $m=-9$.
For Fe$_8$ ($b=0.47$) in zero field one expects a first-order transition but recent measurements of Wernsdorfer \cite{werpriv} suggest that each energy level becomes dominant in the escape at some temperature, i.e., the transition is second order.

Since it is very difficult to find $1/S$ corrections to the quasiclassical results, one has to look for alternative approaches.
For finite spins, the problem can be solved in a purely numerical way, and the calculations can be performed on a modern PC within a reasonable time for $S \leq 100$. 
This is the aim of the present article --- to find out which energy levels make the dominant contribution to the escape rate at different temperatures for different particular cases of the spin model with the Hamiltonian of Eq.\ \ref{sham}.
We show that, in accordance with predictions of quasiclassical model, the dominant level does not necessarily change continuously from the top to the bottom of the metastable well when temperature is lowered.

The rest of the article is organized as follows.
In Sec.\ \ref{sec_basic} we reformulate the theory of thermally assisted tunneling in terms of quantities which can be directly computed for finite spins.
In Sec.\ \ref{sec_uniaxial} we consider the uniaxial model with transverse and longitudinal fields and make a comparison of exact numerical results for the temperature dependence of the tunneling level with earlier perturbative results.
In Sec.\ \ref{sec_biaxial} the calculations are performed for the biaxial model with the field along the hard direction.
We confirm the existence of more complicated scenarios of the escape-rate transitions for this model.

\section{Basic formalism}
\label{sec_basic}

Quasiclassical approach to the quantum statistical metastability considers the spectrum of quantum states as continuous and uses the following expression for the escape rate (see, e.g., Refs.\ \onlinecite{aff81,garmarchu98})
\begin{equation}\label{gamint}
\Gamma \sim \int \!\!dE\; W(E) e^{-(E-E_{\rm min})/T},
\end{equation}
where $W(E)$ is the probability of tunneling at an energy $E$.
The latter can be written as
\begin{equation}\label{we}
W(E) = \frac{1}{1 + \exp[S(E)]} ,
\end{equation}
where for the barriers parabolic near the top the imaginary-time action $S(E)$ goes linearly through zero for $E$ crossing the barrier top level $E=E_c$ and it is {\em analytically continued} into the energy region above the barrier.
In the latter case formula (\ref{we}) describes quantum 
reflections for a particle going over the barrier, with $W(E)$
slightly lower than 1, whereas for the energies below the top of the barrier $W(E)$ is exponentially small in the quasiclassical case.
The action $S(E)$ can be calculated for spin systems with a number of different methods such as the instanton approach, \cite{enzsch86,chugun88,marchu00} mapping on a particle with the WKB approximation, \cite{schwrelvh87,zas90pla,chugar97,garmarchu98,kim99} and the discrete spin WKB method. \cite{lvhsuto86}

At higher temperatures the integral in Eq.\ (\ref{gamint}) is dominated by $E \sim E_c$ which results in the Arrhenius temperature dependence of the escape rate $\Gamma = \Gamma_0 \exp(\Delta U/T)$, where $\Delta U \equiv E_c-E_{\rm min}$.
At lower temperatures the relevant region of energies goes down, which is the regime of thermally assisted tunneling.
Since for quasiclassical systems the crossover between the two regimes occurs at a temperature $T \ll \Delta U$, the integrand in Eq.\ (\ref{gamint}) is a product of two very rapidly increasing and decreasing functions of energy and thus can be approximated by
\begin{equation}\label{gammax}
\Gamma \sim {\rm max}_E \left[W(E) e^{-(E-E_{\rm min})/T}\right].
\end{equation}
Within this approximation, the crossover between the classical and quantum regimes becomes a transition at a well defined temperature $T_0$.
The mathematical description of this transition is analogous to the well known phenomenological Landau model of phase transitions (the Landau theory), as was pointed out in Ref.\ \onlinecite{chugar97}.
The transition can be second or first order.
It should be stressed, however, that it is only a formal analogy and there are certainly no many-body effects in the problem of the escape rates we are studying.
Integration across the maximum of the integrand in Eq.\ (\ref{gamint}) smears the transition and transformes it into a crossover.
In the case of a second-order transition the width of the crossover region around $T_0$ is $\Delta T \propto  1/\sqrt{S}$ (Ref.\ \onlinecite{garmarchu98}) and disappears in the quasiclassical limit.
In the case of a first-order transition there are two competing maxima of the integrand in Eq.\ (\ref{gamint}), and the transition is from one maximum to the other.
In this case the width of the crossover region is even smaller: $\Delta T \propto  1/S$. \cite{garmarchu98} 
We should stress that in spite of the smearing of the escape-rate transition, there is always a fundamental difference between the two situations: one shifting with temperature maximum of $W(E) \exp(-E/T)$ or two competing maxima of $W(E) \exp(-E/T)$.
We will illustrate this difference for various models below.

It is convenient to express the tunneling probability through the quantities which can be directly computed using the quasiclassical formula for the tunnel splitting  \cite{lanlif3}
%
\begin{equation}\label{splt}
\Delta E =
\frac{\omega_E }{\pi}\exp\left[-\frac{S(E)}{2}\right],
\end{equation}
where $\omega_E$ is the frequency of the oscillation in the well at the energy $E$.
Using this formula one obtains
\begin{equation}\label{wesplt}
W(E) = 
\frac{ (\Delta E)^2 }{ (\omega_E/\pi)^2 + (\Delta E)^2 }.
\end{equation}
Since, again, $\Delta E$ given by Eq.\ (\ref{splt}) becomes formally much larger than $\omega_E$ for the energies above the top of the barrier if $S(E)$ is analytically continued into that region, Eq.\ (\ref{wesplt}) gives $W(E)$ fast approaching 1.

Eq.\ (\ref{wesplt}) is the starting point for the numerical solution of the problem for finite spins.
We consider the situation where there are pairs of quasi-degenerate levels in different potential wells and we compute the tunnel splittings for these pairs numerically.
The oscillation frequency $\omega_E$ is nothing else than the difference of the energy of the adjaicent levels in one of the wells: $\omega(E_n)=\delta E_n =E_{n+1}-E_n$.
The discrete analog of Eq.\ (\ref{gamint}) is
\begin{equation}\label{gamsum}
\Gamma \sim \sum_n  \frac{ (\Delta E_n)^2 \delta E_n}
{ (\delta E_n/\pi)^2 + (\Delta E_n)^2 } e^{-(E_n-E_{\rm min})/T}.
\end{equation}
For large spins and low temperatures one has 
\begin{equation}\label{FDef}
\Gamma \sim {\rm max}_n \exp[-F(E_n)/T],
\end{equation}
where
\begin{equation}\label{gamsummax}
\exp\left[-\frac{F(E_n)}T\right] \equiv
\frac{ (\Delta E_n)^2 \delta E_n}
{ (\delta E_n/\pi)^2 + (\Delta E_n)^2 } e^{-(E_n-E_{\rm min})/T}.
\end{equation}
We will call the energy level minimizing the effective free energy $F$ (Ref.\ \onlinecite{chugar97}) and thus maximizing the combined probability of escape the {\em tunneling level} or the level of thermally assisted tunneling.

For the energy levels below the top of the barrier, one has $\Delta E \ll \delta E_n$, whereas above the top of the barrier the levels are not grouped in pairs, i.e., formally, $\Delta E \sim \delta E_n$.
Since for large spins the transition between the two ranges of energy is rather sharp and below the top of the barrier $\Delta E$ changes much faster than $\delta E$, one can look for the maximum of the function \cite{garchu97}
\begin{equation}\label{gamsummax1}
(\Delta E_n)^2 e^{-(E_n-E_{\rm min})/T}.
\end{equation}
Although above the barrier one cannot strictly speak of tunneling, the formula above gives correct results since $\Delta E_n$ becomes weakly dependent on energy and this region is suppressed by the fast decreasing Boltzmann exponential.
In this paper, we will use Eq.\ (\ref{gamsummax}) instead of Eq.\ (\ref{gamsummax1})  since we are going to make a comparison between the exact mumerical solution and the solution that uses the perturbative formula for the level splitings. \cite{gar91jpa}
Since the latter gives $\Delta E \gg \delta E_n$ above the barrier, Eq.\ (\ref{gamsummax}) is more appropriate because it gives physically correct results in this energy range.

\section{Uniaxial model with external field}
\label{sec_field}

This model is the first of spin models for which the first-order escape-rate transition has been found theoretically\cite{garchu97} in the region of small transverse fields $H_x$ using the perturbative formula for the level splittings\cite{gar91jpa,chufri,garchu97}
\begin{eqnarray}\label{SplitPert}
&&
\Delta \varepsilon_{mm'} = \frac{2D}{[(m'-m-1)!]^2}
\nonumber\\
&&
\qquad
\times \sqrt{\frac{ (S+m')! (S-m)! }{ (S-m')! (S+m)! } }
\left( \frac{ H_x }{ 2D } \right)^{m'-m}.
\end{eqnarray}
Here the longitudinal field enters through the resonance condition
\begin{eqnarray}\label{hzres}
&&
E_m = E_{m'}, \qquad m<0, \qquad m' = - m - k 
\nonumber\\
&&
H_z = H_{zk} = kD, \qquad k=0, \pm 1, \pm 2, \ldots
\end{eqnarray}
with $E_m = -Dm^2 - H_z m$, see Fig.\ 1 of Ref.\ \onlinecite{garchu97}. 
Recently corrections to this formula have been ontained in Ref.\ \onlinecite{ulyzas99}.
The use of the perturbative formula for the splittings cannot, however, give an acccurate value of the boundary $h_{xc}\equiv H_{xc}/(2SD)$ between the first- and second-order transitions since the transition occurs at $h_x=h_{xc}=1/4$ (Ref.\ \onlinecite{chugar97}) which is not small.

Our next task is to perform a purely numerical calculation illustrating first- and second-order transitions for finite spins in the transverse field of arbitrary strength.
For $H_x\neq 0$, spin projections on the $z$ axis, $m$, are no longer good quantum numbers.
We will continue, however, to enumerate the exact levels in terms of $m$ to keep a link to the previous work.
At the $k$th resonance, the lowest $k$ levels are not splitted and localized in the right well.
We will formally ascribe them $S_z$ values $m=S, S-1, \ldots, S-k+1$.
Higher levels are grouped in tunnel-splitted pairs which we denote as $\{m,m'\}=\{-S,S-k\}, \{-S+1,S-k-1\}$, etc.
\begin{figure}[t]
\unitlength1cm
\begin{picture}(11,6.5)
\centerline{\epsfig{file=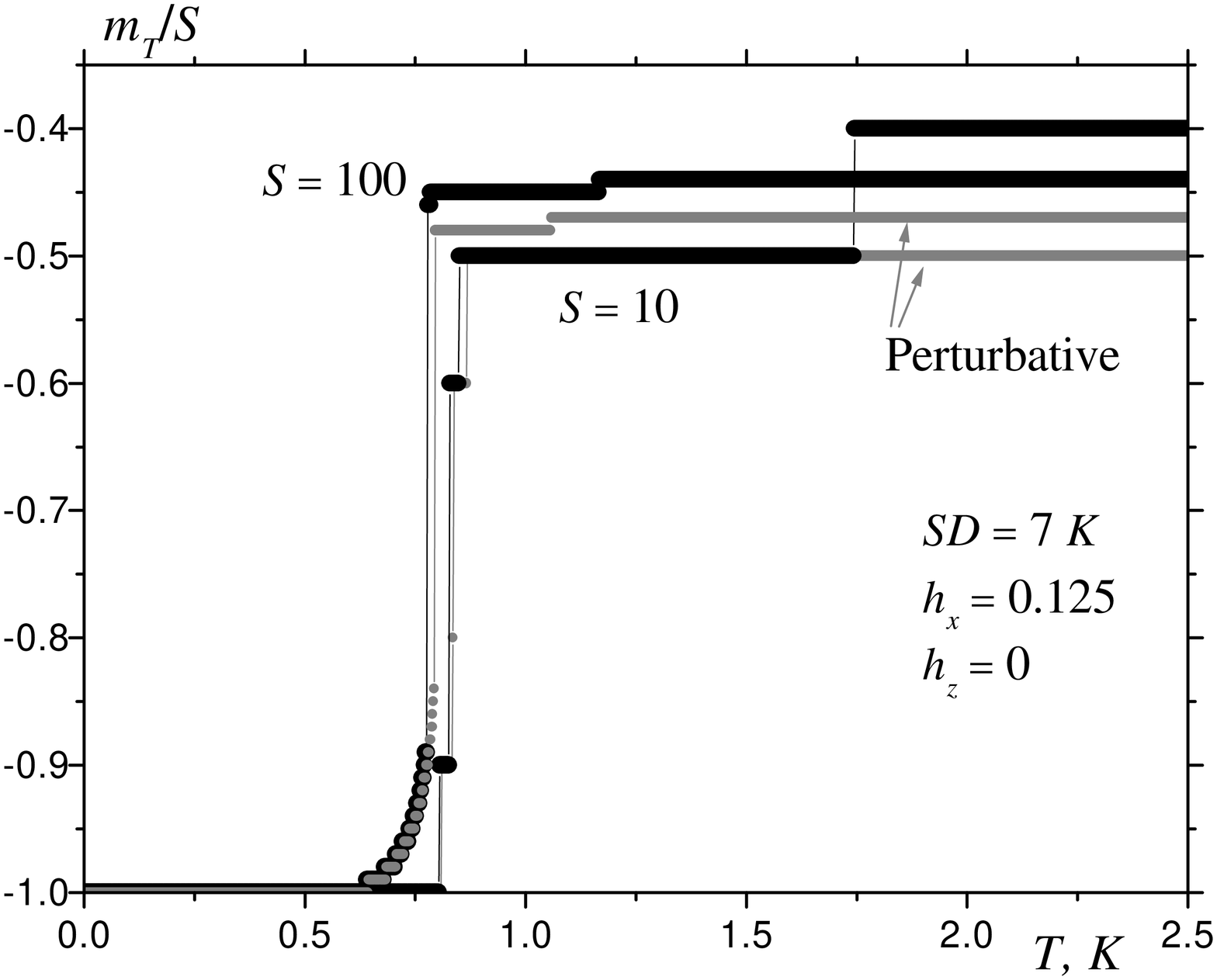,angle=-90,width=9cm}}
\end{picture}
\begin{picture}(11,6.5)
\centerline{\epsfig{file=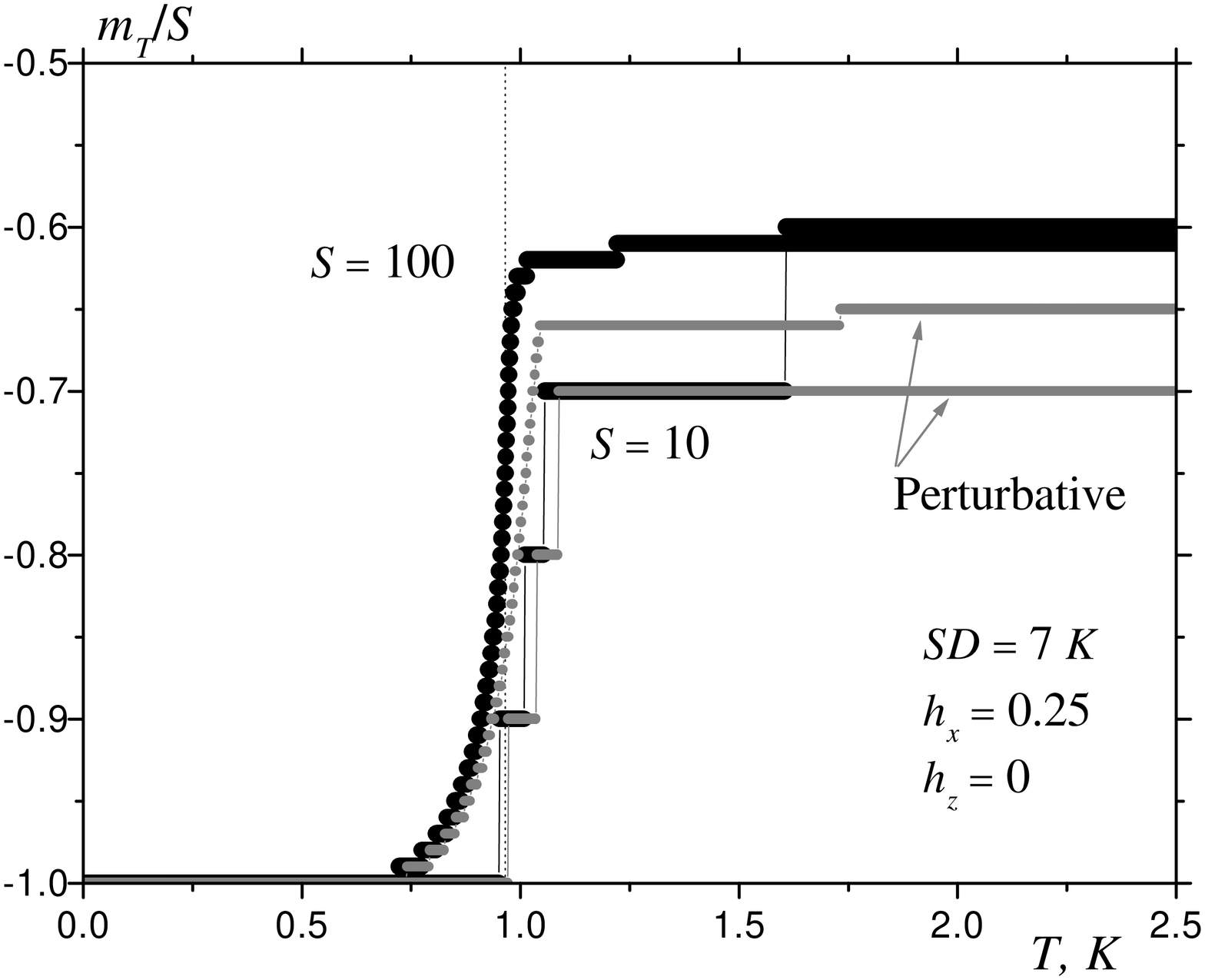,angle=-90,width=9cm}}
\end{picture}
\caption{ \label{qsm_hx_pt} 
Perturbative and exact results for the temperature dependence of the tunneling level $m_T$ for the uniaxial spin model with transverse field.
One can see that the PT holds for small $h_x$ and that finite values of the spin $S$ favor the second-order transition.
}
\end{figure}
\begin{figure}[t]
\unitlength1cm
\begin{picture}(11,6.5)
\centerline{\epsfig{file=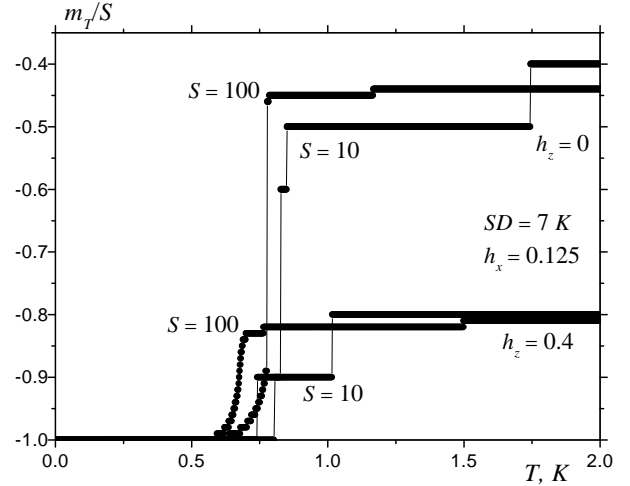,angle=-90,width=9cm}}
\end{picture}
\caption{ \label{qsm_hxhz} 
Temperature dependence of the tunneling level $m_T$ for the uniaxial spin model with transverse and longitudinal field.
}
\end{figure}

For the diagonalization of the spin Hamiltonian we used Wolfram Mathematica which allows one to perform calculations with any desired precision.
We used the parameter set of Mn$_{12}$Ac and ignored the anisotropy of the type $D_4S_z^4$ for simplicity.
Our numerical results for the level splittings reproduce those of Ref.\ \onlinecite{aubflaklaolb96}, where a quantum dimer problem, which is mathematically identical to the spin-in-field problem, has been studied.

The perturbative and exact results for the temperature dependence of the tunneling level $m_T$ which maximizes Eq.\ (\ref{gamsummax}) are shown in Fig.\ \ref{qsm_hx_pt} for $H_z=0$. 
For the field $h_x=0.125$ which can be considered as small, the perturbation theory well describes the transition temperature $T_0$ and the order of transition.
The only noticeable disagreement with the exact results is that regarding the hight of the barrier which is visualized here through the value of $m_T$ in the classical regime.
This is not a surprize since the PT breaks down near the top of the barrier for whatever small $h_x$. \cite{fridiss,garchu97}
For $h_x=0.125$ and $S=100$, many levels are skipped at the transition temperature, thus this transition is first order.
For $h_x=0.125$ and $S=10$, the skipped range is smaller, and the situation is closer to a second-order transition than that for $S=100$.
The dependence on the spin value is even more clearly seen for $h_x=0.25$ which is the exact boundary between first- and second-order transitions in the limit $S\to\infty$.\cite{chugar97}
For $S=10$ there are no skipped levels and the dependence $m_T(T)$ is far from a jump.
For $S=100$ there are no skipped levels, too, but the dependence $m_T(T)$ has a rather high slope near $T_0$.
In the limit $S\to\infty$, the low-slope part of the dependence $m_T(T)$ at $T>T_0$ becomes horizontal, and the derivative $dm_T/dT$ becomes infinite at $T=T_0-0$.
For $S=100$, there are no skipped levels even for $h_x=0.2$.

On Fig.\ \ref{qsm_hxhz} we show exact numerical results for $m_T(T)$ for the $S=10$ and $S=100$ models with $h_x=0.125$ and two values of the longitudinal field, $h_z=0$ and $h_z=0.4$.
These results confirm that increasing of $h_z$ drives the system into the region of the second-order transitions.\cite{garmarchu98}

\begin{figure}[t]
\unitlength1cm
\begin{picture}(11,6.5)
\centerline{\epsfig{file=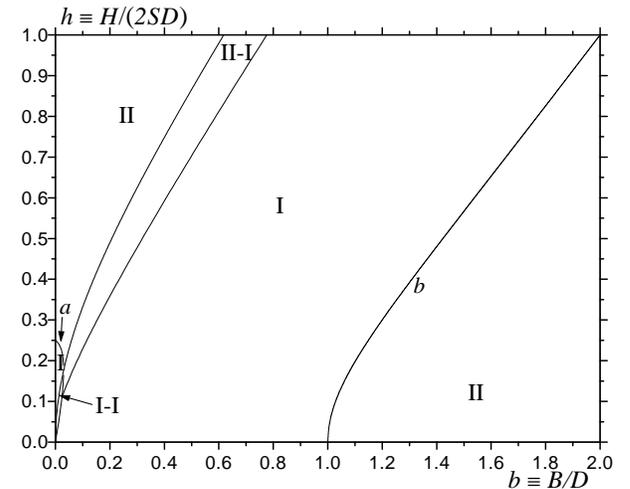,angle=-90,width=9cm}}
\end{picture}
\caption{ \label{qsm_phd} 
For the model ${\cal H} = -DS_z^2 +BS_x^2 - H_xS_x$ phase diagram  includes regions of first- (I) and second-order (II) escape-rate transitions, as well as the regions where a second-order transition is followed by a first-order one (II-I) or the regions of the transitions of the I-I type. \protect\cite{marchu00} 
}
\end{figure}
\begin{figure}[t]
\unitlength1cm
\begin{picture}(11,6.5)
\centerline{\epsfig{file=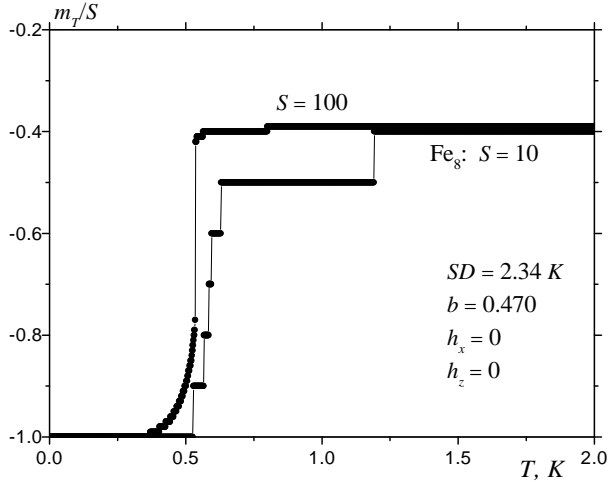,angle=-90,width=9cm}}
\end{picture}
\caption{ \label{qsm_b0} 
Temperature dependence of the tunneling level $m_T$ for the biaxial spin model with $b\equiv B/D=0.470$ in zero field.
Note the second-order transition for $S=10$.
}
\end{figure}
\begin{figure}[t]
\unitlength1cm
\begin{picture}(11,6.5)
\centerline{\epsfig{file=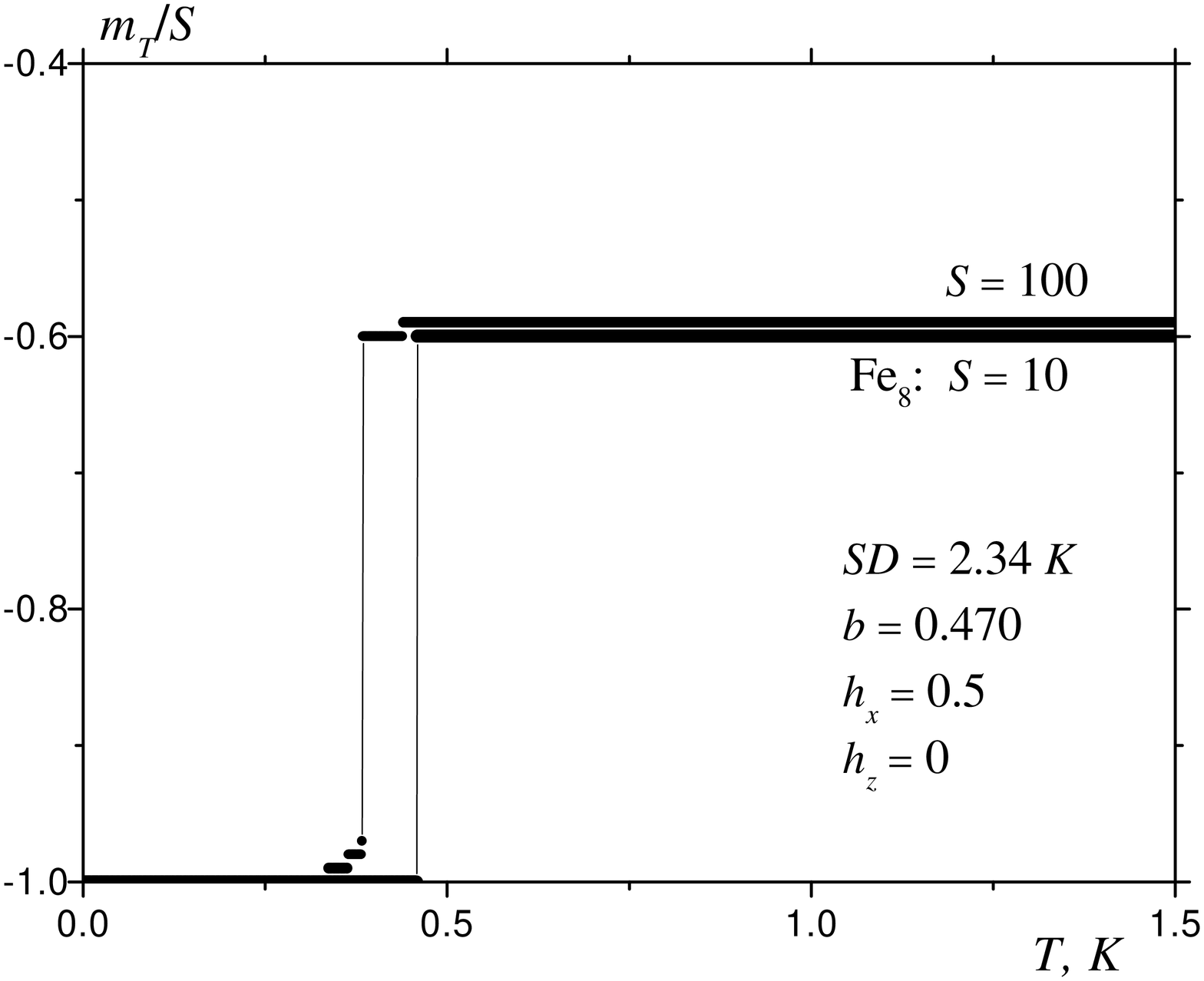,angle=-90,width=9cm}}
\end{picture}
\begin{picture}(11,6.5)
\centerline{\epsfig{file=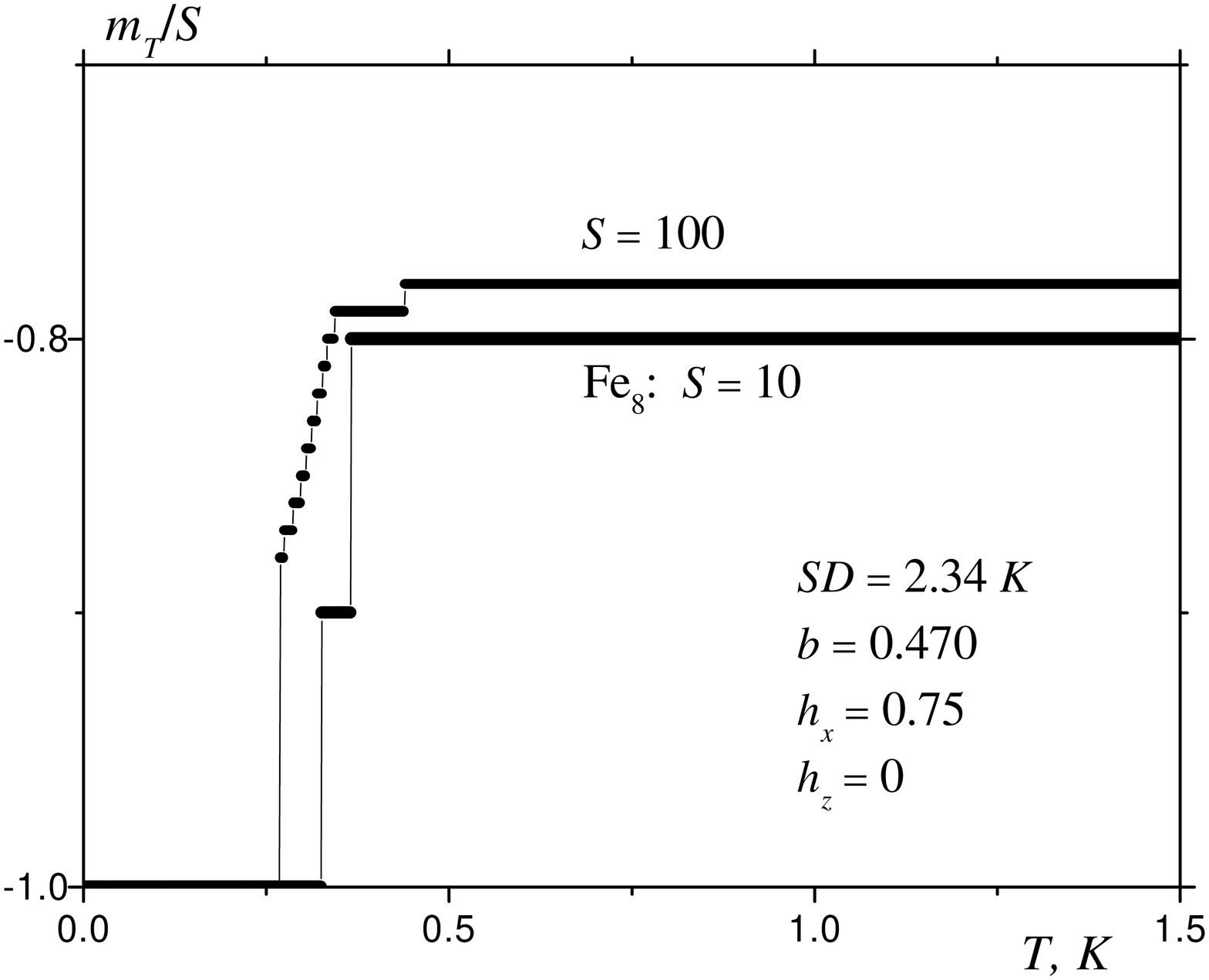,angle=-90,width=9cm}}
\end{picture}
\caption{ \label{qsm_b} 
Temperature dependence of the tunneling level $m_T$ for the biaxial spin model with $b=0.470$ in hard-axis fields $h_x=0.5$ and 0.75.
}
\end{figure}
\begin{figure}[t]
\unitlength1cm
\begin{picture}(11,6.5)
\centerline{\epsfig{file=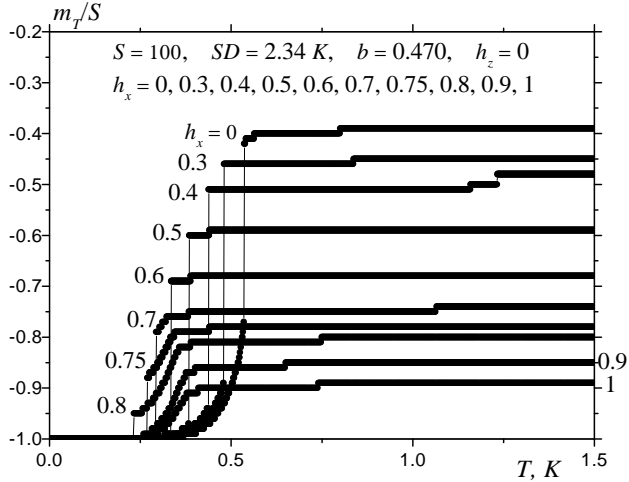,angle=-90,width=9cm}}
\end{picture}
\caption{ \label{qsm_bhx} 
Temperature dependence of the tunneling level $m_T$ for the biaxial spin model with $b=0.470$ and $S=100$ in different hard-axis fields.
}
\end{figure}
\begin{figure}[t]
\unitlength1cm
\begin{picture}(11,6.5)
\centerline{\epsfig{file=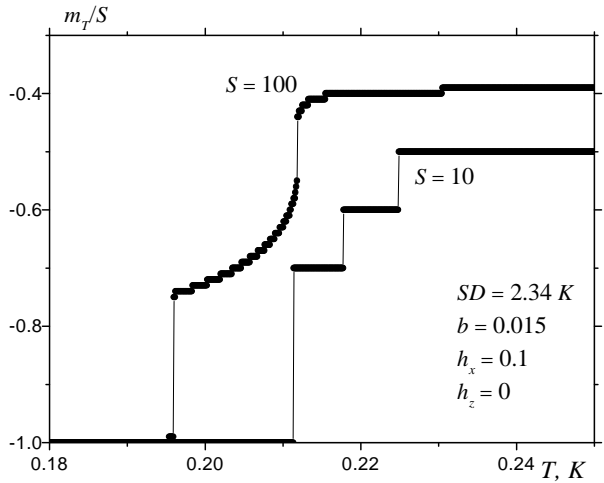,angle=-90,width=9cm}}
\end{picture}
\caption{ \label{qsm_1-1} 
Temperature dependence of the tunneling level $m_T$ for the biaxial spin model with $b=0.015$ and $h_x=0.1$ showing two first-order transitions for $S=100$.
}
\end{figure}
\begin{figure}[t]
\unitlength1cm
\begin{picture}(11,6.5)
\centerline{\epsfig{file=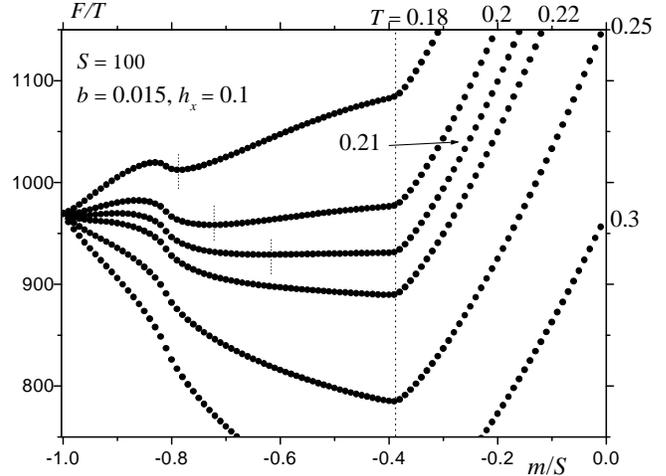,angle=-90,width=9cm}}
\end{picture}
\caption{ \label{qsm_1fvt} 
Effective free energy $F$ of Eq.\ (\protect\ref{FDef}) for $S=100$, $b=0.015$, and $h_x=0.1$ and different temperatures. 
Here a first-order transition is followed by another first-order transition with lowering temperature.
}
\end{figure}
\begin{figure}[t]
\unitlength1cm
\begin{picture}(11,6.5)
\centerline{\epsfig{file=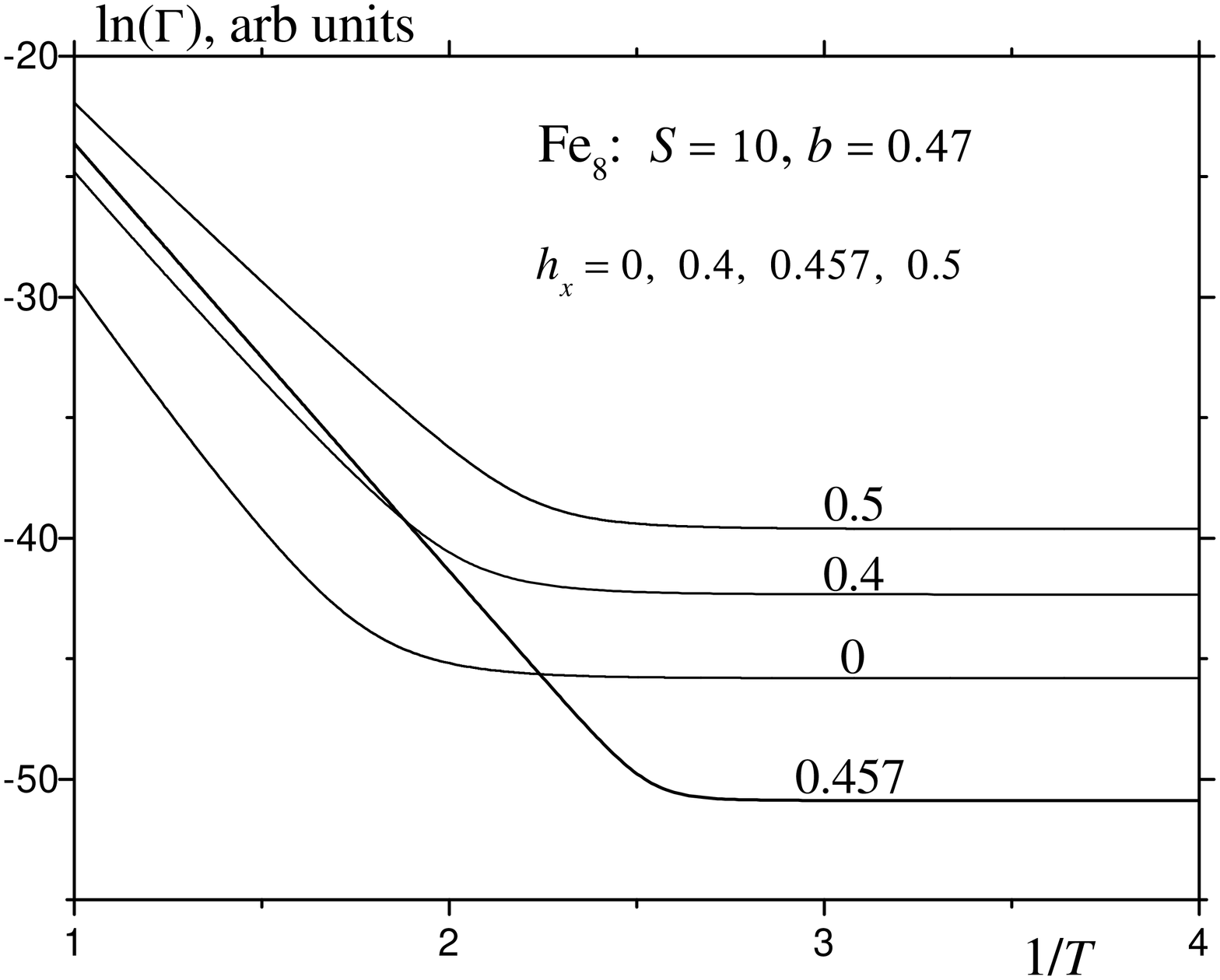,angle=-90,width=9cm}}
\end{picture}
\caption{ \label{qsm_rate} 
The escape rate $\Gamma$ in Fe$_8$ vs $1/T$ for different transverse fields. 
For $h_x=0.475$ tunneling is almost quenched.
}
\end{figure}

\section{Biaxial model}
\label{sec_biaxial}

We will concentrate here on the most interesting model with the field along the hard direction \cite{garg93,werses99,chokim00,marchu00}
\begin{equation}\label{shamhard}
{\cal H} = -DS_z^2 +BS_x^2 - H_xS_x. 
\end{equation}
The phase diagram for the model above, which has been obtained in Ref.\ \onlinecite{marchu00} is shown in Fig.\ \ref{qsm_phd}. 
The boundaries marked $a$ and $b$ have also been obtained in Ref.\  \onlinecite{chokim00}. 
Apart from regions of the first- and second-order transitions marked by I and II, this phase diagram contains the region where a second-order transition is followed by the first-order one (II-I) and a rather narrow region where a first-order transition is followed by another first-order transition (I-I).
The possibility of such multiple transitions has been predicted in Ref.\ \onlinecite{chu92} and here is their first realization in a spin model.

Let us now draw the plots of $m_T(T)$ for $S=10$ and $S=100$ for different transverse fields $h_x$ for the value of the transverse anisotropy $b=0.47$ which is appropriate for Fe$_8$. 
In zero field one expects a first-order transition for $b<1$ in the quasiclassical limit. \cite{liamueparzim98}
This is confirmed by our results for $S=100$ in Fig.\ \ref{qsm_b0}.
However, for $S=10$ there are no skipped levels, although $m_T(T)$ goes rather steep.
This is in accord with recent experiments by Wernsdorfer on Fe$_8$ in zero field, which suggest a second-order transition.\cite{werpriv}

The behavior changes strikingly if a sufficiently strong field $h_x$ is applied.
One can see from the Fig.\ \ref{qsm_b} that for $h_x=0.5$ for both $S=100$ and $S=10$ the transition is the strongest first order.
This effect should be observable on Fe$_8$.
Further increasing the field makes the potential wells so shallow that there are only few levels left.
This makes difficult to make a judgement about the order of the transition for $S=10$.
For $h_x=0.75$ there are no skipped levels for the $S=10$ model and one could speak about a second-order transition.
For $S=100$ one can clearly see a second-order transition followed by a first-order transition, in accordance with the phase diagram on Fig.\ \ref{qsm_phd}.
Dependences $m_T(T)$ for $S=100$ and many different values of $h_x$ are shown in Fig.\ \ref{qsm_bhx}.
The different types of transition in Fig.\ \ref{qsm_bhx} are in accord with the phase diagram of Fig.\ \ref{qsm_phd}.

The most exotic behavior of $m_T(T)$ takes place for small values of transverse anisotropy and field, where one expects two first-order transitions (see Fig.\ \ref{qsm_phd}).
The behavior of $m_T(T)$ for $S=100$, $b=0.015$, and $h_x=0.1$ in Fig.\ \ref{qsm_1-1} confirms the prediction of Ref.\ \onlinecite{marchu00} and shows two jumps.
Note that with lowering temperature $m_T(T)$ for $S=100$ begins to go down continuously and then makes the first jump.
Thus one could speak about the succession of transitions of the type II-I-I, where the second-order transition is solely due to the finite value of the spin and vanishes in the quasiclassical limit.
The same effect also takes place in a simpler uniaxial model with a transverse field $h_x < h_{xc}$.
For $S=10$ the jump at higher temprerature disappears and one thus has a second-order transition followed by a first-order transition.
We illustrate the behavior of the effective free energy $F$ of Eq.\ (\ref{FDef}) for $b=0.015$ and $h_x=0.1$ in Fig.\ \ref{qsm_1fvt}.
(For convenience, we use the value $SD=2.34$~K of Fe$_8$.)
Since the dependence $F(T)$ on the energy level is extremely flat near $T=0.21$~K, the approach using Eq.\ (\ref{FDef}) instead of Eq.\ (\ref{gamsum}) is valid for rather high values of $S$.

As we have mentioned in the Introduction, for the biaxial model with the field along the hard axis and the integer spin, tunneling is quenched whenever\cite{garg93}
\begin{eqnarray}\label{quenching}
&&
H_x = (1+2n)\sqrt{B(B+D)},
\nonumber \\
&&
n = -S, -S+2,\ldots, S-1.
\end{eqnarray}
In the quasiclassical formalism it manifests itself in the vanishing of the prefactor in the tunneling probability.
For $S\gg 1$ the role of the prefactor is difficult to see because the exponential terms dominate.
The rate of thermally assisted tunneling (in the log scale) and thus the transition temperature $T_0$, which depends logarithmically on the prefactor, are significantly reduced only in very close vicinities of quenching points.
For moderate spins such as $S=10$, the quenching effect may be quite substantial.
On Fig.\ \ref{qsm_rate} the tunneling rate and the value of $T_0$ are suppressed near $h_x=0.457$ which corresponds to $n=5$ in Eq.\ (\ref{quenching}).

\section{Discussion}
\label{sec_discussion}

Our direct numerical investigations of quantum-classical escape-rate transitions in spin models with finite $S$ confirmed predictions of quasiclassical approaches in the case of large $S$ and revealed deviations to the favor of a second-order transition for moderate spins.
In particular, in Fe$_8$ in zero field the moderate spin value $S=10$ makes the transition second order.
On the other hand, applying a field along the hard anisotropy axis makes the transition in Fe$_8$ the strongest first order, which can be probably observed in experiment.
For some values of the field tunneling is quenched and the rate of thermally assisted tunneling drops down.

For the biaxial model with the field along the hard axis, we numerically confirmed the existence of the regions where (i) a second-order transition is followed by a first-order transition and (ii) a first-order transition is followed by another first-order transition  with lowering temperature.  \cite{marchu00} 
This model seems to be the only model up to date which demonstrates such a complicated behavior.

This work has been supported by the NSF Grant No.\ DMR-9978882.

\vspace{-0.5cm}

\begin{thebibliography}{10}

\vspace{-1.5cm}

\bibitem[*]{e-gar}
www.mpipks-dresden.mpg.de/$\sim$garanin/, garanin@mpipks-dresden.mpg.de

\bibitem[\dagger]{e-chu}
chudnov@lehman.cuny.edu


\bibitem{chu92}
{E. M. Chudnovsky}, Phys. Rev. A {\bf 46},  8011  (1992).

\bibitem{garchu97}
{D. A. Garanin and E. M. Chudnovsky}, Phys. Rev. B {\bf 56},  11 102  (1997).

\bibitem{chugar97}
{E. M. Chudnovsky and D. A. Garanin}, Phys. Rev. Lett. {\bf 79},  4469  (1997).

\bibitem{garmarchu98}
{D. A. Garanin, X. Mart\'\i nes Hidalgo, and E. M. Chudnovsky}, Phys. Rev. B
  {\bf 57},  13639  (1998).

\bibitem{liamueparzim98}
{J.-Q. Liang, H. J. W. M\"uller-Kirsten, D.-K. Park, and F. Zimmerschied},
  Phys. Rev. Lett. {\bf 81},  216  (1998).

\bibitem{garchu99prb}
{D. A. Garanin and E. M. Chudnovsky}, Phys. Rev. B {\bf 59},  3671  (1999).

\bibitem{zhaetal99}
{Y.-B. Zhang, J.-Q. Liang, H. J. W. M\"uller-Kirsten, S.-P. Kou, X.-B. Wang,
  and F.-C. Pu}, Phys. Rev. B {\bf 60},  12886  (1999).

\bibitem{kim99}
{G.-H. Kim}, Phys. Rev. B {\bf 59},  11847  (1999).

\bibitem{leemueparzim98}
{S.-Y. Lee, H. J. W. M\"uller-Kirsten, D.-K. Park, and F. Zimmerschied}, Phys.
  Rev. B {\bf 58},  5554  (1998).

\bibitem{paryooyoo00}
{C.-S. Park, S.-K. Yoo, and D.-H. Yoon}, Phys. Rev. B {\bf 61},  11618  (2000).

\bibitem{garg93}
{A. Garg}, Europhys. Lett. {\bf 22},  205  (1993).

\bibitem{werses99}
{W. Wernsdorfer and R. Sessoli}, Science {\bf 284},  133  (1999).

\bibitem{chokim00}
{T. Choi and G.-H. Kim}, Physica B {\bf 291},  219  (2000).

\bibitem{marchu00}
{X. Martinez Hidalgo and E. M. Chudnovsky}, J. Phys.: Condensed Matter  (in
  press).

\bibitem{kenetal00}
{A. D. Kent, Y. Zhong, L. Bokacheva, D. Ruiz, D. N. Hendrickson, and M. P.
  Sarachik}, Europhys. Lett. {\bf 49},  521  (2000).

\bibitem{werpriv}
{W. Wernsdorfer},   (private communication).

\bibitem{aff81}
{I. Affleck}, Phys. Rev. Lett. {\bf 46},  388  (1981).

\bibitem{enzsch86}
{M. Enz and R. Schilling}, J. Phys. C {\bf 19},  L711  (1986).

\bibitem{chugun88}
{E. M. Chudnovsky and L. Gunther}, Phys. Rev. Lett. {\bf 60},  661  (1988).

\bibitem{schwrelvh87}
{G. Scharf, W. F. Wreszinski, and J. L. van Hemmen}, J. Phys. A {\bf 20},  4309
   (1987).

\bibitem{zas90pla}
{O. B. Zaslavskii}, Phys. Lett. A {\bf 145},  471  (1990).

\bibitem{lvhsuto86}
{J. L. van Hemmen and A. S\"ut\H o}, Physica B {\bf 141},  37  (1986).

\bibitem{lanlif3}
{L. D. Landau and E. M. Lifshitz}, {\em Quantum {M}echanics} (Pergamon, London,
  1965).

\bibitem{gar91jpa}
{D. A. Garanin}, J. Phys. A {\bf 24},  L61  (1991).

\bibitem{chufri}
{E. M. Chudnovsky and J. R. Friedman},   (unpublished).

\bibitem{ulyzas99}
{V. V. Ulyanov and O. B. Zaslavskii}, Phys. Rev. B {\bf 60},  6212  (1999).

\bibitem{aubflaklaolb96}
{S. Aubry, S. Flach, K. Kladko, and E. Olbrich}, Phys. Rev. Lett. {\bf 76},
  1607  (1996).

\bibitem{fridiss}
{J. R. Friedman}, Ph. D. thesis, City University of New York, 1996  .

\end{thebibliography}

\end{document}